\newif\ifsinglecol
\singlecolfalse 

\ifsinglecol
    \documentclass[journal,12pt,onecolumn,draftclsnofoot]{IEEEtran} 
\else
   \documentclass[conference,a4paper]{IEEEtran}    \usepackage[left=1.43cm,right=1.43cm,top=1.8cm,bottom=4.21cm]{geometry} 
\fi

\usepackage{amsmath,amssymb,amsmath,amsfonts}
\usepackage{bm}
\usepackage{bbm}
\usepackage{graphicx}
\usepackage{cite}
\usepackage{epstopdf}
\usepackage{gensymb}
\usepackage{color}
\usepackage{lipsum}
\usepackage{float}
\usepackage[mathcal]{eucal}
\usepackage{subfiles}
\usepackage[T1]{fontenc}
\usepackage{siunitx}
\usepackage[hyphens]{url}
\usepackage[breaklinks=true]{hyperref}
\usepackage{tabulary}
\usepackage{epsfig,graphics,graphicx,subfig,amssymb,amstext,amsmath,algorithm,algorithmic,wrapfig,multirow}
\usepackage{array}
\usepackage{adjustbox}
\usepackage[dvipsnames]{xcolor}
\usepackage{cite}
\usepackage{times}
\usepackage{placeins}
\usepackage{textcomp}
\usepackage[font=small]{caption}
\usepackage{flushend}
\usepackage{color, colortbl}
\usepackage{tabularx}
\usepackage{bm}
\usepackage{enumerate}
\usepackage{tikz}
\usepackage{pgfplots}
\usetikzlibrary{shapes,arrows}
\usepackage[utf8]{inputenc}
\usepackage{mathtools}
\usetikzlibrary{chains,arrows,calc,positioning}
\usepackage{amssymb}
\usepackage{pifont}
\usepackage{soul}
\usepackage{breqn} 
\usepackage{booktabs} 
\usepackage{multirow}
\usepackage{enumitem}
\usepackage{tabulary}
\usepackage[normalem]{ulem}
\usepackage{dblfloatfix}
\usepackage{makecell}
\usepackage{lipsum}
\usepackage{amsthm}
\usepackage{comment}
\usepackage{filecontents}

\newtheoremstyle{mystyle}
  {}
  {}
  {\itshape}
  {}
  {\bfseries}
  {.}
  { }
  {}
\theoremstyle{mystyle}

\newlength \figwidth
\definecolor{bittersweet}{rgb}{1.0, 0.44, 0.37}
\definecolor{glaucous}{rgb}{0.38, 0.51, 0.71}
\definecolor{gainsboro}{rgb}{0.86, 0.86, 0.86}
\definecolor{babyblueeyes}{rgb}{0.63, 0.79, 0.95}
\definecolor{silver}{rgb}{0.75, 0.75, 0.75}
\definecolor{neoncarrot}{rgb}{1.0, 0.64, 0.26}
\definecolor{Gray}{gray}{0.9}
\definecolor{LightCyan}{rgb}{0.88,1,1}
\definecolor{BackgroundLightBlue}{rgb}{0.97,0.97,1}
\definecolor{BackgroundGray}{gray}{0.98}

\newcommand{\black}[1]{\textcolor{black}{#1}}
\newcommand{\blue}[1]{\textcolor{black}{#1}}

\makeatletter

 \let\oldforeign@language\foreign@language
 \DeclareRobustCommand{\foreign@language}[1]{%
   \lowercase{\oldforeign@language{#1}}}

\def\nb0{{\mathbf{0}}}
\def\nb1{{\mathbf{1}}}

\def\nbbE{{\mathbb{E}}}

\makeatother

\IEEEoverridecommandlockouts

\begin{document}

\bstctlcite{IEEEexample:BSTcontrol}

\title{Spatially Consistent Air-to-Ground Channel Modeling via Generative Neural Networks}

\ifsinglecol
   {\author{
\IEEEauthorblockN{Amedeo Giuliani, Rasoul Nikbakht, Giovanni Geraci, Seongjoon Kang,\\Angel Lozano, and Sundeep Rangan
}
\thanks{A. Giuliani and R. Nikbakht are with Centre Tecnol\`{o}gic de Telecomunicacions de Catalunya (CTTC), Spain. G. Geraci is with Telef\'{o}nica and UPF, Spain. S. Kang and S. Rangan are with NYU Wireless, USA. A. Lozano is with UPF, Spain. This work was supported by grants PID2021-123999OB-I00, CEX2021-001195-M, TSI-063000-2021-56/57, the UPF-Fractus Chair, and ICREA. Remcom contributed by providing the Wireless Insite software.}
}}%
\else
    {\author{
\IEEEauthorblockN{Amedeo Giuliani, Rasoul Nikbakht, Giovanni Geraci, Seongjoon Kang, Angel Lozano, and Sundeep Rangan
}
\thanks{A. Giuliani and R. Nikbakht are with Centre Tecnol\`{o}gic de Telecomunicacions de Catalunya (CTTC), Spain. G. Geraci is with Telef\'{o}nica and UPF, Spain. S. Kang and S. Rangan are with NYU Wireless, USA. A. Lozano is with UPF, Spain. This work was supported by grants PID2021-123999OB-I00, CEX2021-001195-M, TSI-063000-2021-56/57, the UPF-Fractus Chair, and ICREA. Remcom contributed by providing the Wireless Insite software.}
}}
\fi

\maketitle

\begin{abstract}
This article proposes a generative neural network architecture for spatially consistent air-to-ground channel modeling. 
The approach considers the trajectories of uncrewed aerial vehicles
along typical urban paths, capturing spatial dependencies within received signal strength (RSS) sequences from multiple cellular base stations (gNBs). 
Through the incorporation of conditioning data, the model accurately discriminates between gNBs and drives the correlation matrix distance between real and generated sequences to minimal values. 
This enables evaluating performance and mobility management metrics
with spatially (and by extension temporally) consistent RSS values, rather than independent snapshots. For some tasks underpinned by these metrics, say handovers, consistency is essential. 
\end{abstract}

\section{Introduction}

Next-generation mobile networks are envisioned to reliably connect uncrewed aerial vehicles (UAVs) \cite{GerGarAza2022,wu20205g,ZenGuvZha2020,KarGerJaf2023}. This will require a re-engineering of existing deployments, e.g., via multiantenna techniques, dedicated infrastructure, and cellular-satellite integration \cite{GarGerLop2019,ChoGuvSaa2021,GerLopBen2022}. Accurate channel models are crucial for evaluating these and other solutions. 

Extending statistical channel models to air-to-ground scenarios is challenging due to the complex dependencies on UAV altitude, orientation, and building height, among other aspects \cite{3GPP36777,KhaGuvMat2019,KhuCheZha2018}. 
Data-driven approaches have been put forth for site-specific channel modeling, mapping spatial locations to channel parameters through regression \cite{huang2018big,ostlin2010macrocell,dall2011channel}. Generative neural networks (GNNs) provide an alternative for non-site-specific propagation modeling \cite{XiaRanMez2020a,8663987,8685573}. However, existing works can only produce
independent channel snapshots, 
failing to capture how signals fluctuate during motion. While adequate for performance evaluations where a marginal distribution suffices, say for coverage, this is insufficient and might cause artifacts 
when designing networks to cope with UAV mobility.

This paper proposes a new GNN architecture for spatially consistent air-to-ground channel modeling. 
\black{
Specifically, the large-scale channel behavior is represented, as embodied by the local-average received signal strength (RSS); this subsumes every aspect save for the small-scale fading. 
In fact, the transmit power is set to $0$ dBm and the antennas are taken to be omnidirectional, such that the RSS (in dBm) exactly equals the large-scale channel gain (in dB).
}
The approach considers a UAV flying along random
typical trajectories in an urban area.
To capture dependencies within sequences RSS from multiple base stations (gNBs) on buildings of varying heights, a generative adversarial network (GAN) is introduced that incorporates two types of conditioning data: the distance sequence from a specific gNB, and the gNB index. This allows generating RSS sequences along designated trajectories corresponding to a particular gNB. Ray tracing is used for data provisioning, although the approach is compatible with measured data.  This model:
\begin{itemize}
    \item
    Successfully learns the marginal distribution of RSS from different gNBs and accurately discriminates between them, despite their similarity.    
    \item
    Generates spatially consistent RSS sequences, with the correlation matrix distance between real and generated sequences converging to minimal values. Data augmentation through sequence self-convolution further enhances the accuracy of the model.
\end{itemize}

Once trained, the model can serve as the workhorse of system-level evaluations consisting of:
\begin{itemize}
    \item Producing UAV trajectories and gNB locations stochastically, based on a deployment model providing conditioning data for each UAV-gNB link.
    \item Sampling random vectors from a prior distribution and feeding them to the model, along with the conditioning data, thus obtaining a sequence of RSS values.
\end{itemize}

\black{
With a faster spatial sampling rate, a similar approach could be employed to incorporate the small-scale fading and
produce the full multipath channel response, including path gains, delays, and angles.
}
\section{Problem Formulation}\label{sect:sys_mod}

A UAV and several gNBs are considered, respectively acting as receiver and transmitters. By reciprocity, their roles are interchangeable. On a circumscribed area in the city of Boston, the gNBs located on buildings of varying heights and the UAV follows random typical trajectories at a fixed height of $30$ m. Fig.~\ref{fig:traj_example} depicts a typical such trajectory.

\subsubsection*{Spatially consistent generative model}
Each gNB-UAV link is characterized by the RSS.
%
%
The collection of RSS values for the $k$th trajectory are denoted by
\begin{equation}
    \boldsymbol{x}_i^k = \left[ p_1, \ldots, p_N \right],
\end{equation}
where $i$ is the gNB identifier, $N$ the number of spatial steps, and $p_n$ the RSS value at the $n$th step. 
Similarly, the evolution of the UAV-gNB distance is denoted by
\begin{equation}
    \boldsymbol{u}_i^k = \left[ d_1, \ldots, d_N \right],
\end{equation}
where $d_n$ is the 3D distance at the $n$th step. 
In the sequel, we employ a conditioning variable $c$ equal to the gNB identifier $i$, yet the methodology can be extended to more general conditioning variables, such as the height and gNB type.
The goal is to capture dependencies within the RSS sequences for multiple specific gNBs across a set of typical trajectories, i.e., to model the conditional distribution $p(\boldsymbol{x} \, | \, \boldsymbol{u}, c)$. This can be achieved by a generative model described by the mapping
\begin{equation}
    \boldsymbol{x} = f(\boldsymbol{z}, \boldsymbol{u}, c),
\end{equation}
where $\boldsymbol{z}$ is a random latent vector following a prior distribution $p(\boldsymbol{z})$, usually uniform or Gaussian. This generating function $f(\boldsymbol{z}, \boldsymbol{u}, c)$ is to be trained with data. This formulation allows differentiation among gNBs while capturing similarities in the RSS spatial distribution.

\subsubsection*{Exploitation of the model}
Once trained, this generative model can be conveniently applied in simulations. UAV trajectories and gNB locations can be stochastically generated based on a deployment model, providing the condition vector $\boldsymbol{u}$ for each UAV-gNB link. Random vectors
$\boldsymbol{z}$ can be sampled from the prior distribution, and with $\boldsymbol{u}$, $\boldsymbol{z}$, and the categorical information $c$, the sequence of RSS values $\boldsymbol{x}$ can be obtained. These RSS values can be generated for intended and interfering links, enabling the computation of many quantities of interest as experienced by a UAV along its route, say signal-to-interference-plus-noise ratios, bit rates, and frequency and success of handovers (cell reselections).

\begin{figure}[!t]
    \centering
    \includegraphics[width=\linewidth]{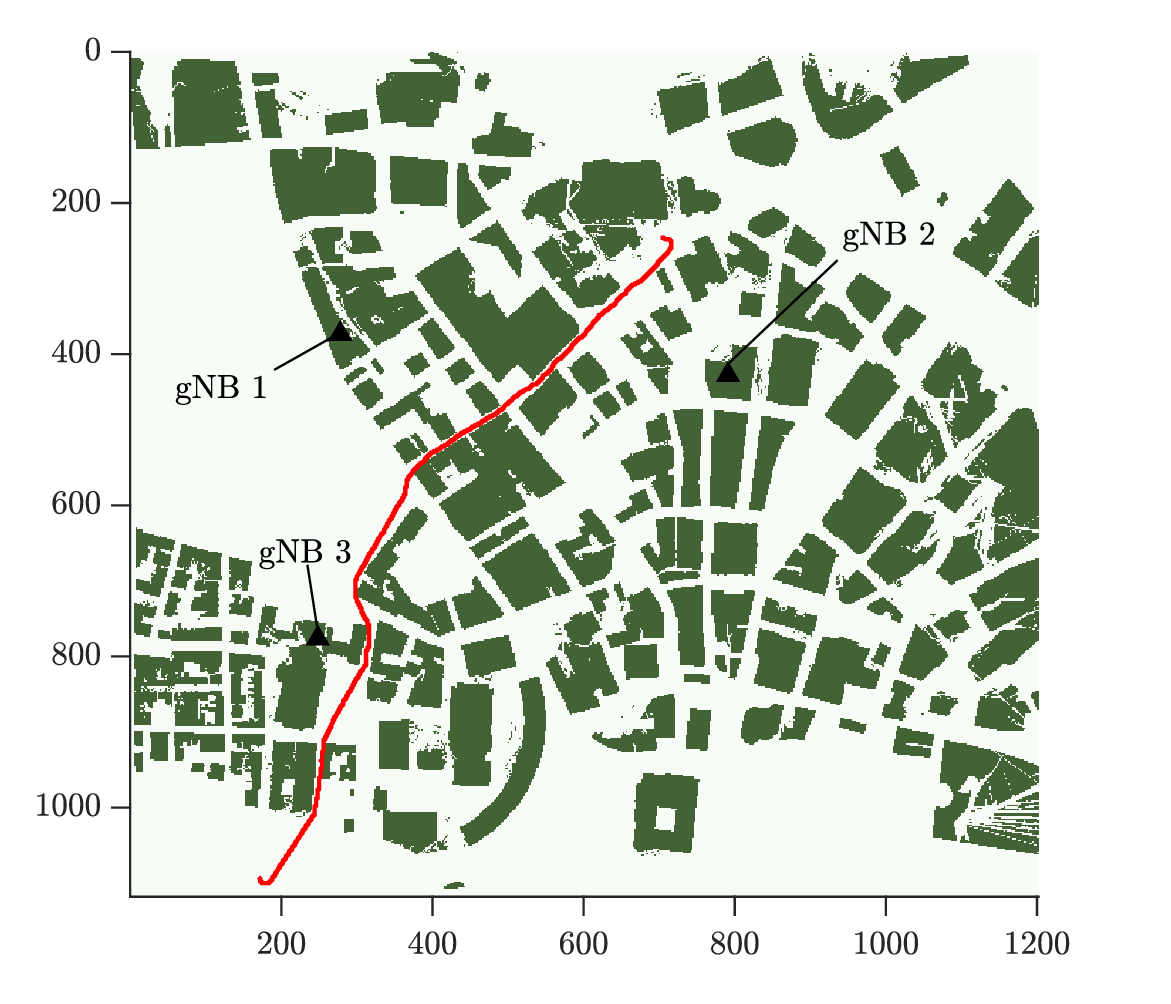}
    \caption{Simulation area showing an example of a UAV trajectory (red) between buildings (green) and the locations of three gNBs ($\triangle$).}
    \label{fig:traj_example}
\end{figure}

\section{Methodology}\label{sect:method1}

\subsection{Generative Model Architecture}

\begin{figure}
    \centering
    \includegraphics[width=\linewidth]{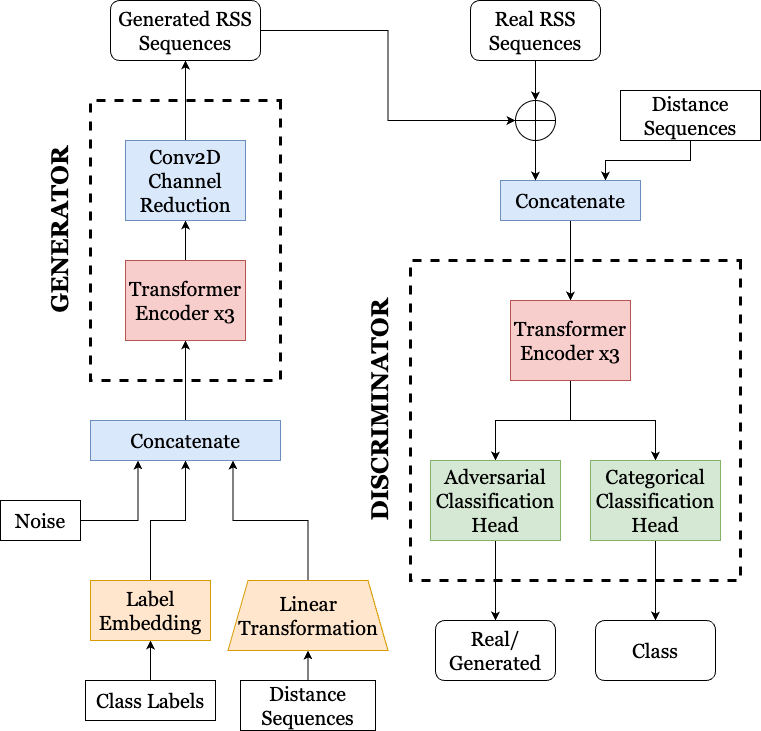}
    \caption{Proposed generative channel model architecture.}
    \label{fig:arch_multiple}
\end{figure}

The proposed architecture incorporates two types of conditioning data, namely the distance sequences $\boldsymbol{u}$ and the gNB indices $c$. 
This enables the generation of RSS sequences along specific trajectories for a particular gNB. 
The architecture, depicted in Fig.~\ref{fig:arch_multiple}, is based on the transformer time-series conditional GAN (TTS-CGAN) \cite{li2022tts} and multivariate time-series conditional GAN (MTS-CGAN) \cite{madane2022transformer} with appropriate modifications. 
Specifically, the conditioning on gNB indices follows the auxiliary classifier GAN paradigm of TTS-CGANs while that on distance sequences relies on the classical conditional GAN paradigm of MTS-CGANs. Both the generator ($G$) and the discriminator ($D$) share a similar structure. They consist of three \texttt{TransformerEncoder} layers, each with five attention heads. Each layer includes a multi-head attention module followed by a feed-forward multi-layer perceptron (MLP) with a Gaussian error linear unit (GELU) activation function. Normalization layers precede both blocks, and dropout layers (with drop rate $\rho = 0.5$) follow them. Generator and discriminator only differ in their input and output layers. Their structure is summarized in Table~\ref{table:parameters}.

\subsubsection*{Generator}

The input consists of a label embedding for the conditioning variable $c$ and a linear transformation of the distance sequence $\boldsymbol{u}$, to allow for their concatenation together with the random latent vector $\boldsymbol{z}$. The concatenated vector is then mapped to a sequence of the same length with 50 embedding dimensions.
With this configuration, the task is to generate RSS sequences related to a specific gNB, based on the distance sequences experienced from that gNB. 

\subsubsection*{Discriminator}

The input module includes a patch and a positional embedding layer. Letting $C$ be the number of gNBs, a sequence of $N$ steps, whether real or generated, can be viewed as an image of shape $(C, 1, W)$, i.e., with height $1$, width $W=N$, and $C$ color channels.
This image is evenly divided into $W/P$ patches, each with shape $(C, 1, P)$, and a learned positional encoding value is added to each patch to preserve its positional information. 

The output module consists of a binary classification layer to classify signals as true or generated, and a multi-class classification layer to determine the originating gNB. Based on the distance sequences provided as conditioning data, the discriminator has two objectives:
\begin{itemize}
\item Adversarial classification, correctly classifying time series as true or generated. \item Categorical classification, accurately assigning the label to the input series.
\end{itemize}

\begin{table}
\centering
\caption{Generator, discriminator, and training parameters.}
\label{table:parameters}
\def\arraystretch{1.2}
\begin{tabulary}{\columnwidth}{|C|C|} \hline
\textbf{Generator layers} & \textbf{Number of weights} \\ \hline
Distance encoder (Linear + LeakyReLU) & $12850$ \\ \hline
Embedding & $386560$ \\ \hline
\texttt{TransformerEncoder} $\times 3$ & $1330 \times 3$ \\ \hline
Channel reduction (Conv2D) & $11$ \\ \hline\hline
\textbf{Discriminator layers} & \textbf{Number of weights} \\ \hline
Patch Embedding & $2550$ \\ \hline
\texttt{TransformerEncoder} $\times 3$ & $30650 \times 3$ \\ \hline
Adversarial classification head & $151$ \\ \hline\hline
\textbf{Training parameter} & \textbf{Value} \\ \hline
Generator and discriminator learning rates & $3 \cdot 10^{-4}$ and $10^{-4}$ \\ \hline
$\beta_1$ and $\beta_2$ & $0.9$ and $0.999$ \\ \hline
Batch size and patch size & $64$ and $16$ \\ \hline
Number of iterations & $5 \cdot 10^5$ \\ \hline
\end{tabulary}
\end{table}

\subsection{Loss Function}

The game between discriminator and generator spans two levels: adversarial classification and categorical classification. 
Let $\hat{\boldsymbol{x}} = G(\boldsymbol{z} \mid \boldsymbol{u}, c)$ be the generator's output while $D_{\text{adv}}(\cdot)$ is the output of the discriminator's adversarial head and $D_{\text{cls}}(\cdot)$ the output of the discriminator's classification head.

\subsubsection*{Adversarial Classification} 

The least-squares (LS) loss function is adopted \cite{mao2017least}, with the discriminator $D$ minimizing 

\ifsinglecol
   {\begin{equation}\label{eq:d_adv_loss}
    \begin{aligned}
        L_{\mathrm{LS}}^{(D)} &= \frac{1}{2} \, \nbbE_{X \sim p_{X}(\boldsymbol{x})} \! \left[(D_{\text{adv}}(\boldsymbol{x} \mid \boldsymbol{u}) - t_{\textrm{R}})^2\right] + \frac{1}{2} \, \nbbE_{Z \sim p_{Z}(\boldsymbol{z})} \! \left[(D_{\text{adv}}(\hat{\boldsymbol{x}} \mid \boldsymbol{u}) - t_{\textrm{F}})^2\right],
    \end{aligned}
\end{equation}}%
\else
    {\begin{equation}\label{eq:d_adv_loss}
    \begin{aligned}
        L_{\mathrm{LS}}^{(D)} &= \frac{1}{2} \, \nbbE_{X \sim p_{X}(\boldsymbol{x})} \! \left[(D_{\text{adv}}(\boldsymbol{x} \mid \boldsymbol{u}) - t_{\textrm{R}})^2\right] \\
        & \quad + \frac{1}{2} \, \nbbE_{Z \sim p_{Z}(\boldsymbol{z})} \! \left[(D_{\text{adv}}(\hat{\boldsymbol{x}} \mid \boldsymbol{u}) - t_{\textrm{F}})^2\right],
    \end{aligned}
\end{equation}}%
\fi

\noindent where $t_{\textrm{R}}$ and $t_{\textrm{F}}$ are the binary labels (targets) for real and generated data, respectively. The generator $G$ minimizes
\begin{equation}\label{eq:g_adv_loss}
    L_{\mathrm{LS}}^{(G)} = \frac{1}{2} \nbbE_{Z \sim p_{Z}(\boldsymbol{z})} [(D_{\text{adv}}(\hat{\boldsymbol{x}} \mid \boldsymbol{u}) - t_{\textrm{R}})^2],
\end{equation}
thus attempting to make $D$ classify generated data $\hat{\boldsymbol{x}}$ as real.

\subsubsection*{Categorical Classification}

Based on the cross-entropy (CE) loss, discriminator and generator respectively minimize
\begin{align}\label{eq:d_cls_loss}
    L_{\mathrm{CE}}^{(D)} & = - \nbbE[\log D_{\text{cls}}(\boldsymbol{x} \mid \boldsymbol{u})] \\
\label{eq:g_cls_loss}
    L_{\mathrm{CE}}^{(G)} & = - \nbbE[\log D_{\text{cls}}(\hat{\boldsymbol{x}} \mid \boldsymbol{u})].
\end{align}

\subsubsection*{Overall Loss}
Discriminator and generator respectively seek the minimum of the total loss functions
\begin{align}\label{eq:d_total_loss}
    L_{\rm D} & = L_{\mathrm{LS}}^{(D)} + L_{\mathrm{CE}}^{(D)} \\
\label{eq:g_total_loss}
    L_{\rm G} & = L_{\mathrm{LS}}^{(G)} + L_{\mathrm{CE}}^{(G)}.
\end{align}

\subsection{Validation through First- and Second-Order Statistics}


To evaluate the model's ability to capture the underlying probability distribution, the first- and second-order statistics of real and generated signals are compared.
To that end, the marginal cumulative distribution function (CDF) of the RSS and the correlation matrix distance (CMD) between real and generated RSS sequences are examined.  

\subsubsection*{Marginal Distribution}

\black{For a given gNB, consider two sets
containing the real and generated RSS sequences, both sets having shape $(B, W)$ with $B$ and $W$ being the batch size and sequence length, respectively.}
Each set is flattened into a 1D vector of length $n = B \cdot W$ over which the CDF is computed.

\subsubsection*{Correlation Matrix Distance}

The sequence length $W$ is regarded as the number of random variables while $B$ is treated as the number of realizations.
Covariance matrices $\boldsymbol{\Sigma}_1$ and $\boldsymbol{\Sigma}_2$ are then computed for the sets of real and generated sequences,
%
%
and from those the correlation matrices: from $\boldsymbol{\Sigma}_1$, we construct $\boldsymbol{A}_1 = \sqrt{\text{diag}(\boldsymbol{\Sigma}_1)}$ and then normalize $\boldsymbol{\Sigma}_1$ into
\begin{equation}
    \boldsymbol{R}_1 = \boldsymbol{A}_1^{-1}\,\boldsymbol{\Sigma}_1\,\boldsymbol{A}_1^{-1}.
\end{equation}
Similarly, from $\boldsymbol{\Sigma}_2$ we construct $\boldsymbol{A}_2$ and obtain $\boldsymbol{R}_2$.
\black{Matrices $\boldsymbol{R}_1$ and $\boldsymbol{R}_2$ contain every second-order statistic for the real and generated RSS sequences, respectively, with no assumptions on stationarity. (If wide-sense stationarity holds, the matrices are Toeplitz, but in general that is not the case for the RSS.)}
The CMD between $\boldsymbol{R}_1$ and $\boldsymbol{R}_2$ is \cite{1543265}
\begin{equation}
    \text{CMD}(\boldsymbol{R}_1, \boldsymbol{R}_2) = 1 - \frac{\text{trace}(\boldsymbol{R}_1 \,\boldsymbol{R}_2)}{\|\boldsymbol{R}_1\|_{\textrm{F}} \, \|\boldsymbol{R}_2\|_{\textrm{F}}} \in [0,1],
\end{equation}
where $\|\cdot\|_{\textrm{F}}$ denotes Frobenius norm. The CMD tends to $1$ if $\boldsymbol{R}_1$ and $\boldsymbol{R}_2$ are maximally different, while it vanishes if they are equal up to a scaling factor, which is the desired outcome.

\section{Numerical Results}

Next, the dataset production process is detailed and two case studies are presented to showcase the model's effectiveness.

\subsection{Dataset Production}

Due to the limited availability of data on UAV channels, the ray tracing package Wireless InSite by Remcom is employed
\black{and the RSS is computed by adding the ray powers at any given location \cite[Sec. 4.2]{rappaport2010wireless}.}

\subsubsection*{Deployment Scenario}

A 3D representation of a region measuring $1200$ m $\times$ $1120$ m is imported, corresponding to the city of Boston (see Fig.~\ref{fig:traj_example}). The representation includes terrain and building data. Transmitting gNBs are manually positioned on three rooftops, $30$ m above street level. These sites are potential locations for providing connectivity to UAVs or other aerial devices \cite{GerLopBen2022,KanMezLoz2021}. 

\subsubsection*{Ray Tracing}
Simulations are conducted at $28$ GHz, which is the dominant frequency for emerging 5G mmWave systems.
Buildings are modeled as made of concrete with permittivity $5.31$ F/m and conductivity of $0.024$ S/m, and the maximum number of reflections and diffractions are set to 6 and 1, respectively. For each gNB, an \black{RSS map} is generated over the entire region, sampled every $2$ m at a height of $30$ m, \black{and assuming 0\,dBm transmit power and unitary antenna gains at both transmitter and receiver. The $2$-m sample spacing is a conservative choice that ensures minimal change in the large-scale channel behavior across consecutive samples. (Further modeling the scall-scale fading would require a smaller spacing, on the order of half the wavelength, as that is the minimum coherence distance of the small-scale process.)
}

\subsubsection*{RSS Sequence Production}

UAVs are placed along multiple random trajectories at $30$ m of height, produced via Matlab's Vehicle Network Toolbox, each consisting of $600$--$800$ steps with a $2$-m interval. 
The RSS sequences $\boldsymbol{x}$ are created by recording the RSS values at each intersection between a trajectory and the power map associated with each gNB. The 3D distance between each intersection and the gNB's location are computed to construct the distance sequences $\boldsymbol{u}$. Let $\boldsymbol{X} = [\boldsymbol{x}^{(1)}, \ldots, \boldsymbol{x}^{(K)}]$ and $\boldsymbol{U} = [\boldsymbol{u}^{(1)}, \ldots, \boldsymbol{u}^{(K)}]$ assemble all RSS sequences and all 3D distances, where $K$ is the total number of each. 
Further, with $c$ identifying the gNB assigned to each sequence, let $\boldsymbol{c} = [c_1, \ldots, c_K]$. 
Thus, $\boldsymbol{X}$, $\boldsymbol{U}$, and $\boldsymbol{c}$ encapsulate all of the information gathered during the data production process.

\subsubsection*{Training Data Augmentation} 

Augmentation is employed to further expand the training data.
This facilitates the model's convergence, particularly in cases where data is scarce (say in the case of a single gNB, as discussed next).
While this step may not be necessary for ray tracing data, as more trajectories and more gNB power maps could be produced, it becomes crucial when the model is trained with measurements, which are considerably more time-consuming. 
Using the Python package \texttt{tsaug}, 
a self-convolution operation is applied to each RSS sequence with a flat kernel window of size $20$. The resulting convolved sequences are then appended to the existing dataset.


\subsection{Case Study I: Single gNB}
\label{Sec:Numerical_single}

To begin with, let us consider only gNB~2 in Fig.~\ref{fig:traj_example}. Without the need to differentiate among different gNBs, the categorical classification can be disabled and the training can focus solely on the classical adversarial game between generator and discriminator. This entails updating the weights of the discriminator and generator based only on (\ref{eq:d_adv_loss}) and (\ref{eq:g_adv_loss}). 

\begin{figure}[!t]
    \centering
    \includegraphics[width=\linewidth]{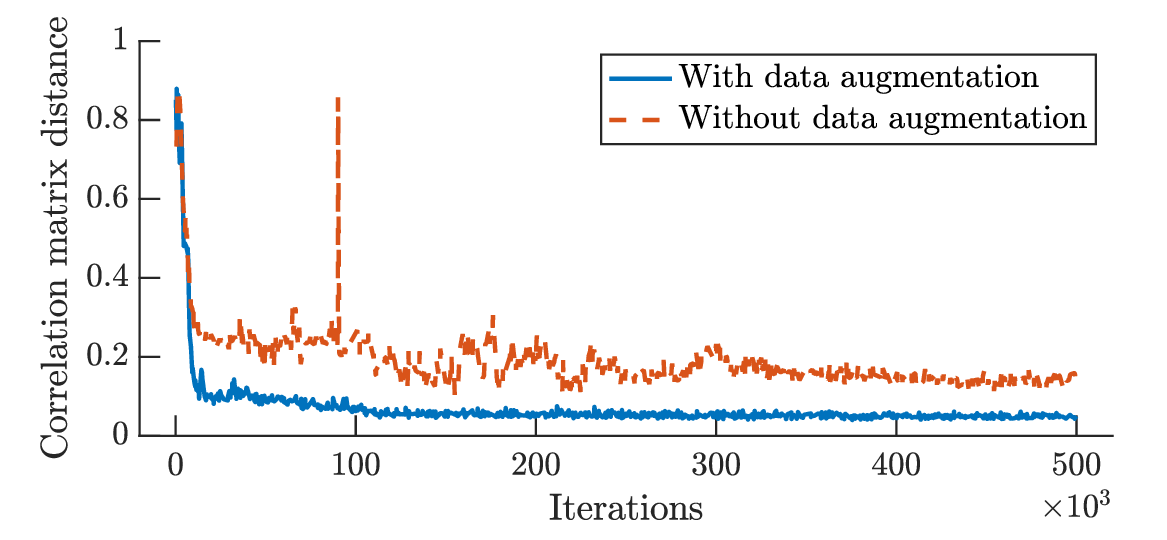}
    \caption{CMD computed on the test set vs. number of training iterations for the case of a single gNB, with and without data augmentation.}
    \label{fig:cmd_comp_1}
\end{figure}

Fig.~\ref{fig:cmd_comp_1} displays the CMD on the test set as a function of the number of training iterations, with and without training data augmentation via convolution. Data augmentation does help to achieve a smaller correlation matrix distance between real and generated RSS sequences, resulting in a more accurate and spatially consistent model. The final value of the CMD on the test set is reported in Table~\ref{table:cmd}, providing further evidence of the model's accuracy.

Fig.~\ref{fig:corrmat_comp} illustrates the evolution of the correlation matrix for the generated RSS sequences as the training progresses, 
when data augmentation is employed during training, with the conditioning distance sequences extracted from the test set. Note that the rows/columns of Fig.~\ref{fig:corrmat_comp} can be interpreted as the autocorrelation of the RSS sequences. The correlation matrix for the corresponding real RSS sequences is also displayed, demonstrating the convergence. 

\ifsinglecol
   {\begin{figure}[!t]
    \centering
    \subfloat[Generated at iteration 10,000]{
        \includegraphics[trim={1.9cm 2.1cm 2.2cm 1.5cm},clip,width=0.5\linewidth]{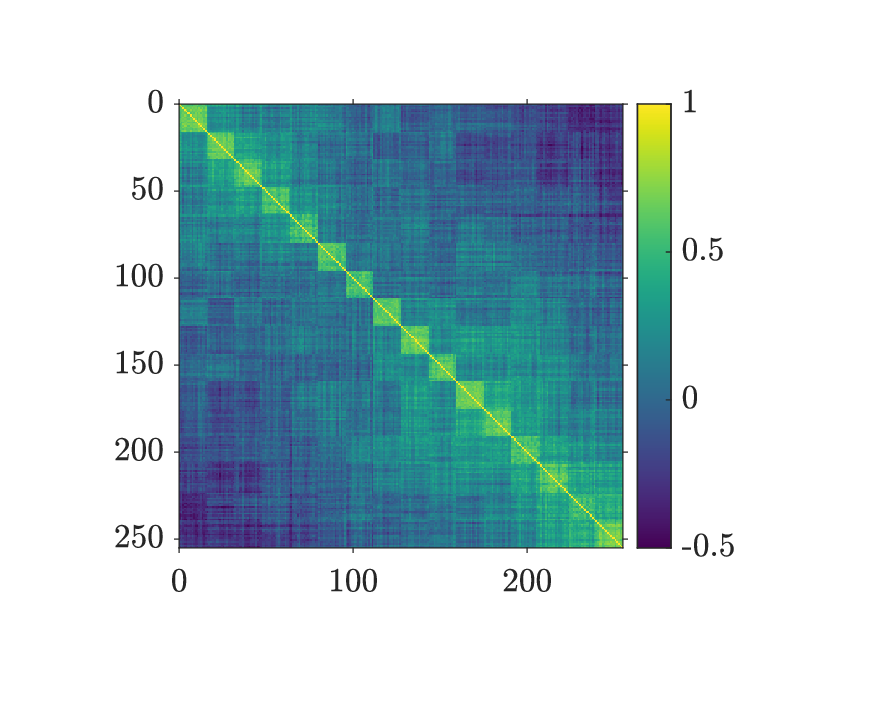}
        \label{fig:corrmat_comp_c}
    }
    \subfloat[Generated at iteration 20,000]{
        \includegraphics[trim={1.9cm 2.1cm 2.2cm 1.5cm},clip,width=0.5\linewidth]{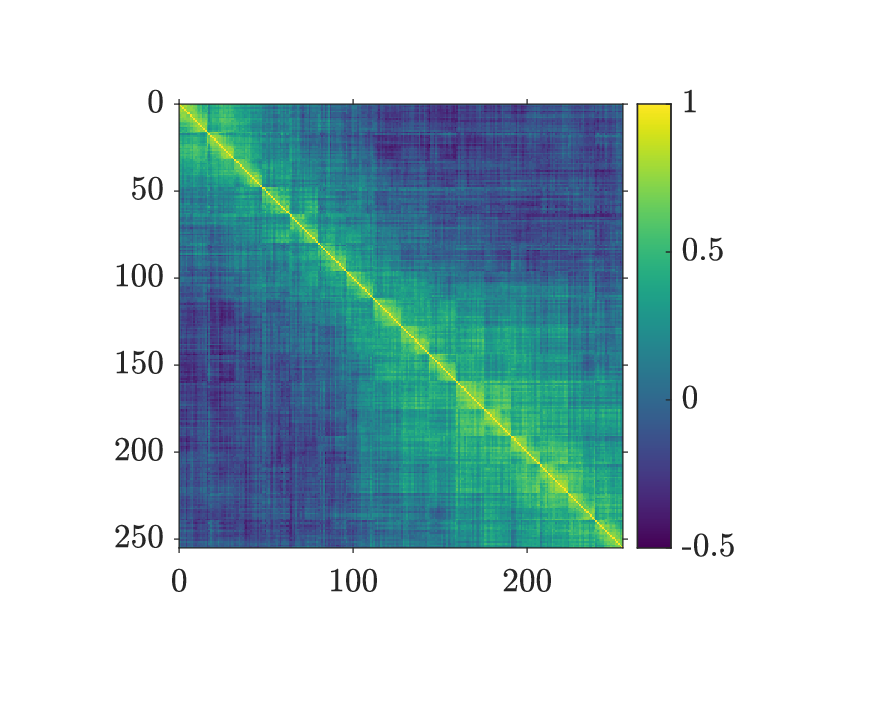}
        \label{fig:corrmat_comp_d}
    }\\
    \subfloat[Generated at iteration 100,000]{
        \includegraphics[trim={1.9cm 2.1cm 2.2cm 1.5cm},clip,width=0.5\linewidth]{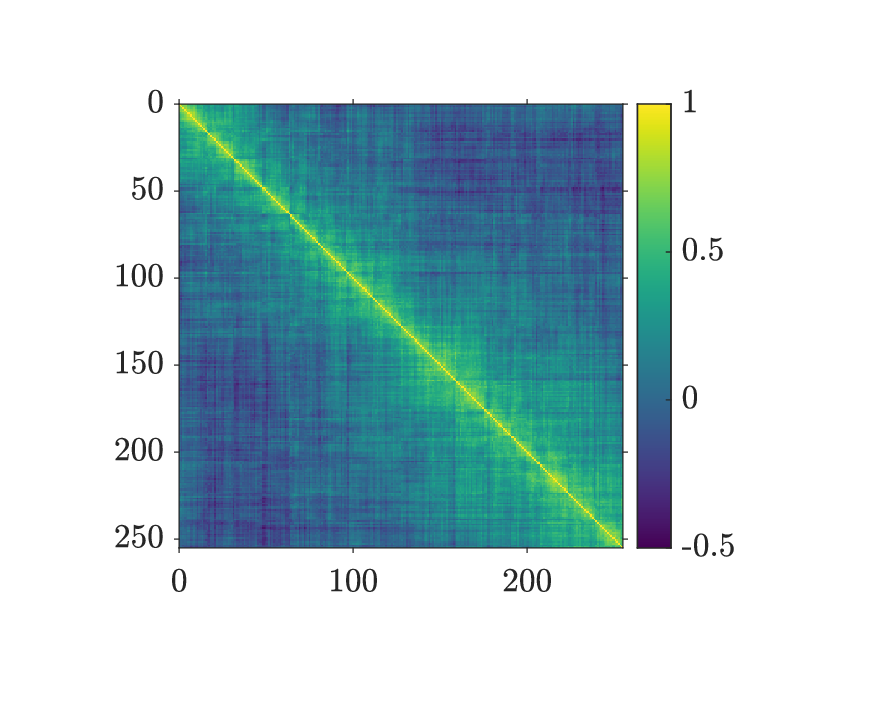}
        \label{fig:corrmat_comp_g}
    }
    \subfloat[Real]{
        \includegraphics[trim={1.9cm 2.1cm 2.2cm 1.5cm},clip,width=0.5\linewidth]{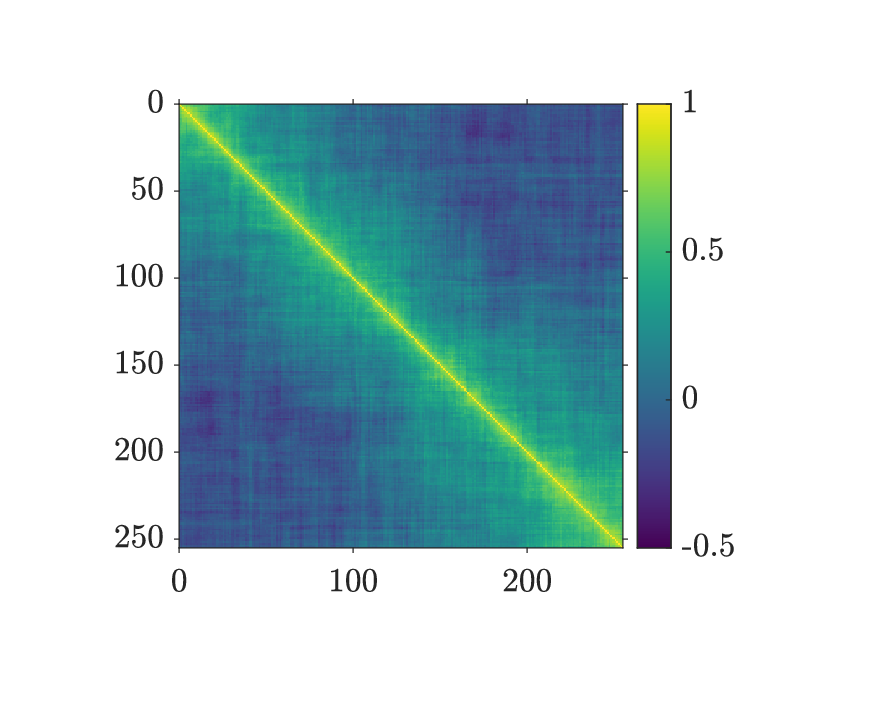}
        \label{fig:corrmat_comp_a}
    }
    \caption{Evolution of the correlation matrix for generated RSS sequences as compared to the one of real RSS sequences.}
    \label{fig:corrmat_comp}
\end{figure}}%
\else
    {\begin{figure*}[!t]
    \centering
    \subfloat[Generated at iteration 10,000]{
        \includegraphics[trim={1.9cm 2.1cm 2.2cm 1.5cm},clip,width=0.45\figwidth]{Figures/corr_fake_10000.eps}
        \label{fig:corrmat_comp_c}
    }
    \subfloat[Generated at iteration 20,000]{
        \includegraphics[trim={1.9cm 2.1cm 2.2cm 1.5cm},clip,width=0.45\figwidth]{Figures/corr_fake_20000.eps}
        \label{fig:corrmat_comp_d}
    }
    \subfloat[Generated at iteration 100,000]{
        \includegraphics[trim={1.9cm 2.1cm 2.2cm 1.5cm},clip,width=0.45\figwidth]{Figures/corr_fake_100000.eps}
        \label{fig:corrmat_comp_g}
    }
    \subfloat[Real]{
        \includegraphics[trim={1.9cm 2.1cm 2.2cm 1.5cm},clip,width=0.45\figwidth]{Figures/corr_real_0.eps}
        \label{fig:corrmat_comp_a}
    }
    \caption{Evolution of the correlation matrix for generated RSS sequences as compared to the one of real RSS sequences.}
    \label{fig:corrmat_comp}
\end{figure*}}%
\fi

\begin{table}
\centering
\caption{Final CMD values with and without data augmentation.}
\label{table:cmd}
\def\arraystretch{1.2}
\begin{tabulary}{\linewidth}{|l|C|C|} \hline
\textbf{Augmentation} & \textbf{CMD with one gNB} & \textbf{CMD with three gNBs} \\ \hline
No & $0.1609$ & $0.1025$ \\ \hline
Yes & $0.0661$ & $0.0874$ \\ \hline
\end{tabulary}
\end{table}
    

\subsection{Case Study II: Multiple gNBs}
\label{Sec:Numerical_multiple}

Next, let us evaluate the generator's ability to capture different distributions by reintroducing the categorical classification head of the discriminator. The model is trained with data from the three gNBs in Fig.~\ref{fig:traj_example}, updating the discriminator and generator according to  (\ref{eq:d_total_loss}) and (\ref{eq:g_total_loss}).

\begin{figure}
    \centering
    \includegraphics[width=0.9\linewidth]{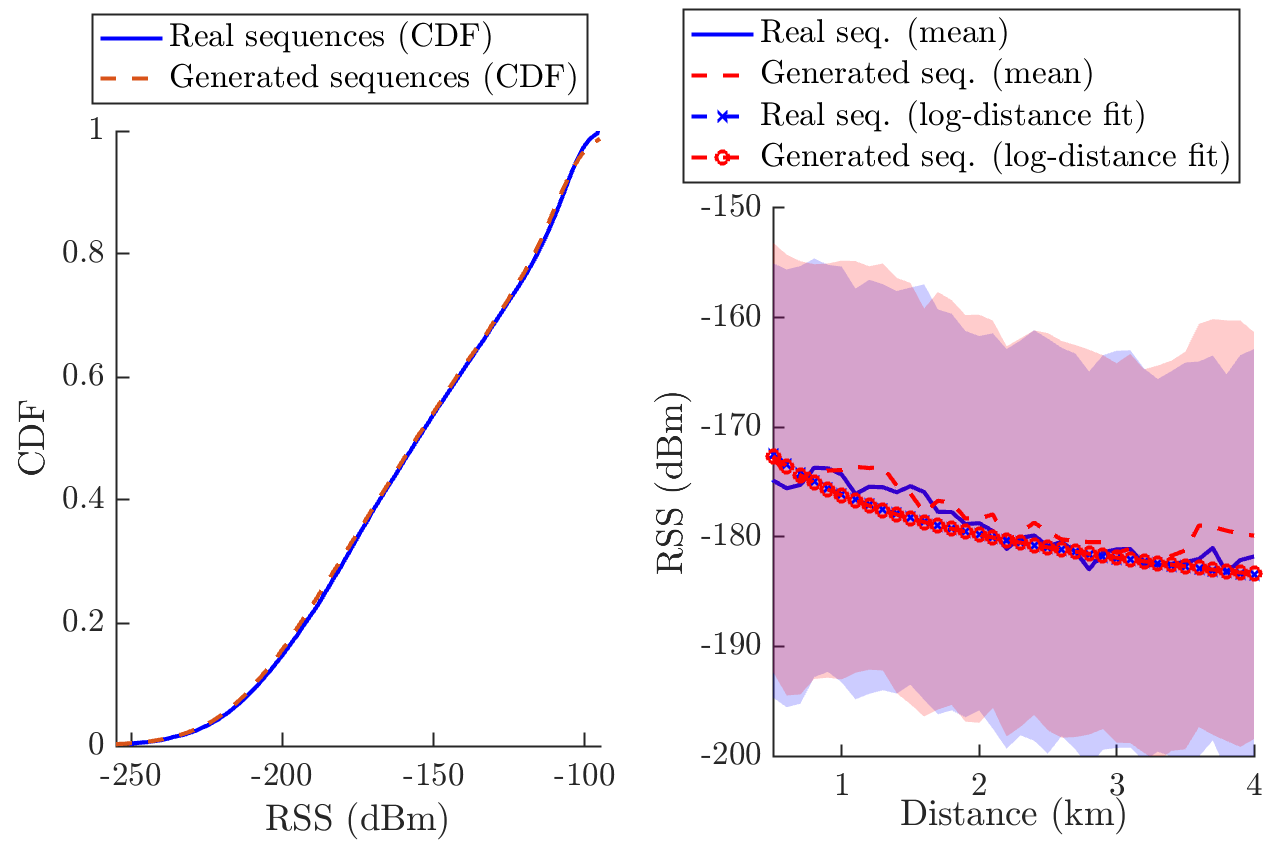}
    \caption{\blue{Left-hand side: CDF of the real and generated RSS values. Right-hand side: mean (solid and dashed lines) plus/minus standard deviation (shaded areas) of the real and generated RSS vs. distance and respective log-distance least-squares fits. Generated sequences are obtained by training with data augmentation.}}
    \label{fig:CDF_meanStd_1gnb_horizontal}
\end{figure}

Fig.~\ref{fig:CDF_meanStd_1gnb_horizontal} presents, on the left-hand side, the CDF of real and generated RSS values corresponding to the distances in the test set, for gNB\,2 with data augmentation. On the right-hand side, it presents 
the mean (solid and dashed lines) and standard deviation (shaded areas) of the real and generated RSS as a function of the UAV-gNB distance. \blue{Log-distance least-squares fits are also shown, with the reference
distance set to the central point of the axis (2.25 km) and
path loss exponents of 2.38 (real) and 2.44 (generated).}
The model is seen to successfully learn the distribution of the RSS and its dependence on the distance.

\begin{figure}[!t]
    \centering
    \includegraphics[width=0.95\linewidth]{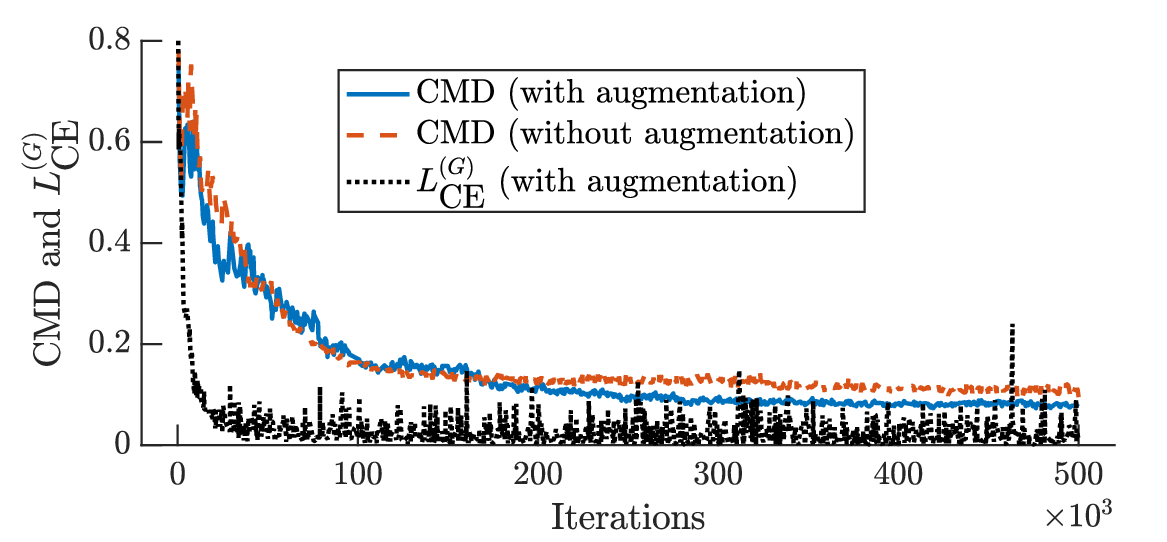}
    \caption{CMD averaged across multiple gNBs vs. training iterations. The categorical classification loss of the generator is also shown.}
    \label{fig:cmd_comp_3}
\end{figure}

Fig.~\ref{fig:cmd_comp_3} depicts the CMD on the test set, averaged for the three gNBs, as a function of the number of training iterations, with and without data augmentation. As the CMD decreases, the model learns to generate spatially consistent RSS sequences that are stochastically similar to the real ones. Data augmentation aids in achieving an even smaller CMD between real and generated sequences. 
The final values of the correlation matrix distance are reported in Table \ref{table:cmd}. 
Fig.~\ref{fig:cmd_comp_3} also displays the evolution of the classification loss (\ref{eq:g_cls_loss}) at the generator for the case of data augmentation.%
\footnote{The discriminator classification loss (\ref{eq:d_cls_loss}), not shown, follows a similar trend.}
As the total loss nears zero, the model successfully discriminates among gNBs.

\section{Conclusion}

This paper has introduced a GNN architecture for spatially consistent air-to-ground channel modeling. The approach effectively captures the spatial dependencies in RSS sequences \black{(equivalently in large-scale channel gains)} from multiple gNBs and can be instrumental for system-level evaluations of various metrics. 
While the presented case studies relied on ray tracing data, the model can also be trained with field measurements.

\black{
Follow-up work could aim---at the expense of a faster spatial sampling rate and increased complexity---at incorporating the small-scale fading,
altogether delivering the full multipath response with path gains, delays, and angles of arrival and departure for all propagation paths. These would further extend the applicability of the model to multiantenna communication. Exploring the model's ability to generalize across environments would be another relevant research direction.
}


\bibliographystyle{IEEEtran}
\bibliography{journalAbbreviations, main}

\end{document}